# Russian Troll Account Classification with Twitter and Facebook Data

Dominic Lewinski

Md Rashidul Hasan

University of New Mexico,

Department of Mathematics and Statistics

## Abstract

In this analysis, we work with the data set that was compiled by Darren Linvill and Patrick Warren, along with a representative sample of Facebook ads that were released by the House Intelligence Committee Minority. The goal of this analysis is to use the categories defined by Linvill and Warren in the Twitter data and investigate if these categories exist in Facebook ads. This begin to give us insights to the tactics used between the two social media services. Further, we try to replicate Linvill and Warren's original categorization of the Twitter data. Lastly, we investigate what categories may exist in the Facebook data.

## Introduction

In the years leading up to and following the 2016 presidential election, Russia and the Internet Research Agency (IRA) attempted to influence the elections outcome by creating chaos and spreading disinformation. The IRA took to social media services such as Twitter and Facebook to sow discord online. These social media services allowed the IRA to create, publish and ultimately expose American citizens to propaganda. In a joint statement by the Department of Homeland Security and the Office of the Director of National Intelligence, the departments determined that senior Russian government officials had directed an e-mail hacking-and-dumping campaign to interfere with the election. Further documentation of Russia's involvement comes from the Intelligence Community Assessment (ICA). According to the ICA, "Russia's goals were to undermine public faith in the US democratic process, denigrate Secretary Clinton, and harm her electability and potential presidency". Further investigation into Russia's interference during the election led to Special Counsel Robert Mueller indicting 13 Russian individuals and three Russian organizations. The House Intelligence Committee Minority has held open hearing with social media companies and released Twitter and Facebook content for researchers to analyze. The extent of their influence on the election may never be known.

**Data Sets**

The Twitter data set was compiled by Darren Linvill and Patrick Warren, both professors at Clemson University, using Clemson's Social Media Listening Center. The data set contains approximately three million tweets from over 2,800 IRA affiliated Twitter handles. Dates range from 2012 up to 2018. There is also additional information about each tweet, such as, number of followers, if the tweet was a retweet, publish date and time of tweet and account category. The account category consists of eight categories: Commercial, Fearmonger, Hashtag Gamer, Left Troll, News Feed, Non-English, Right Troll and Unknown. Where Left and Right Troll are tweeting about left and right winged politics. News Feed is imitating local news. The previous three make up the bulk of the set. Hashtag Gamer is used for distracting users by having users participate in "hashtag games". Fearmonger is used to spread news about fake crisis. Fearmonger appears the least in the data set.

The Facebook data set is a small representative sample of Facebook advertisements released by the House Intelligence Committee Minority. The full data set consists of approximately 80,000 pieces of organic content from 470 IRA-created accounts. It is estimated that nearly 126 million Americans were exposed to the 80,000 pieces of organic content. The Committee Minority plans to make the full data set public in the future as it is not available currently. The publicly available Facebook ads consist of approximately 3,500 ads. Each ad was published in a separate pdf file which contains the ads metrics, content and image of the ad. An example of an ad can be found in the appendix under Image A. To protect individual's identity, serval ads have redacted information or images.

## Data Cleaning and Exploration

The Twitter data set did not require a large amount of preprocessing. Of the 21 variables in the set, the main interest was in: ‘content’, ‘publish_date’, ‘account_category’ and ‘tco1_step1’. From content, hashtags and users tweeted at were extracted because it was theorized that these could give better predictive power on their own. Content also contained many URLs, emoticon and various punctuations. These were all removed along with capitalization to reduce variability among the same words. Content that contained two or less words were removed. This is done because some of the algorithms perform weakly with low word count. The publish_date variable was reformatted and all tweets that fell below March 2015 or above November 2017 were removed to match the Facebook timeline. The tco1_step1 variable contained information about retweets and that information was extracted. For this analysis, Commercial, Non-English and Unknown were dropped from the set. The decision to drop these categories is due to the nature of the Facebook data. The Facebook data does not contain Non-English ads or commercial advertisement.

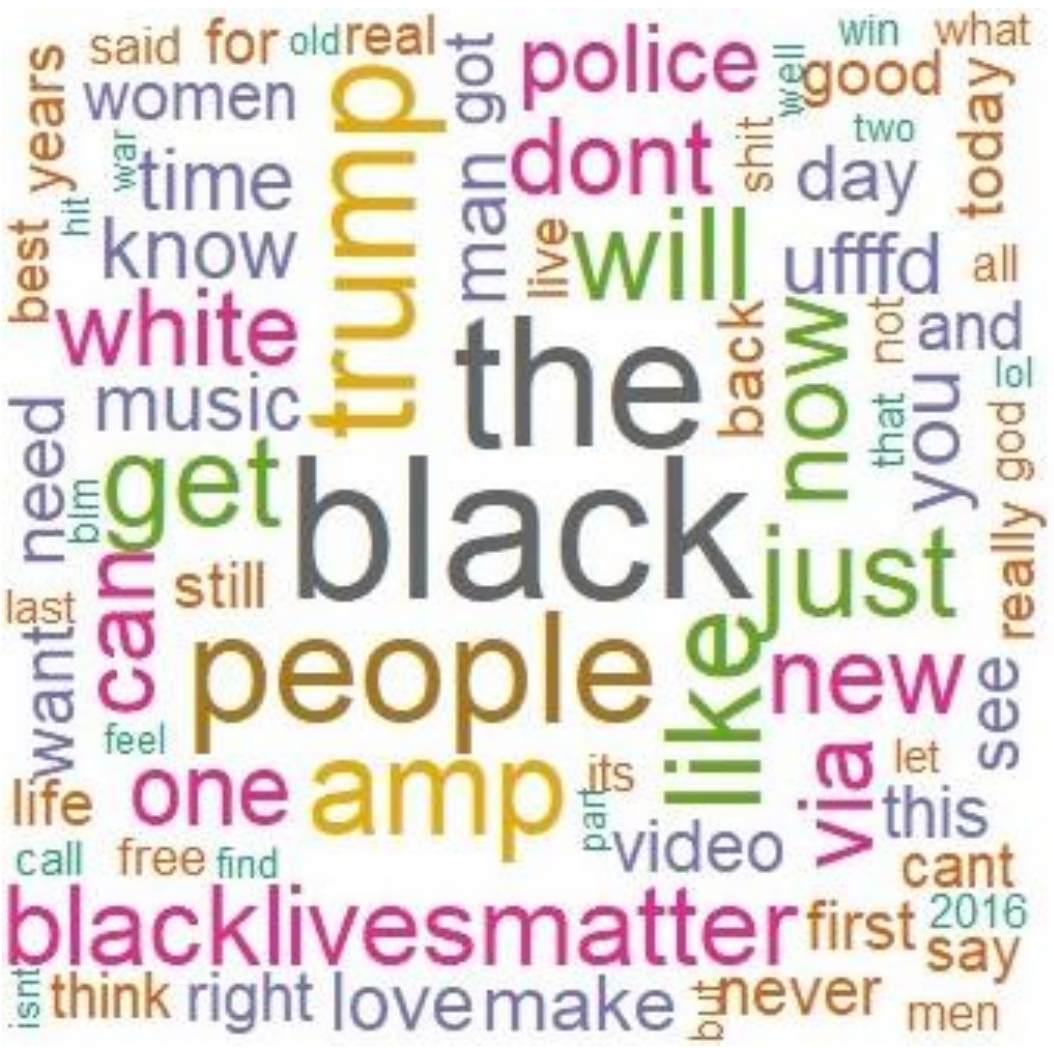

Image 1: Left Troll most frequent words

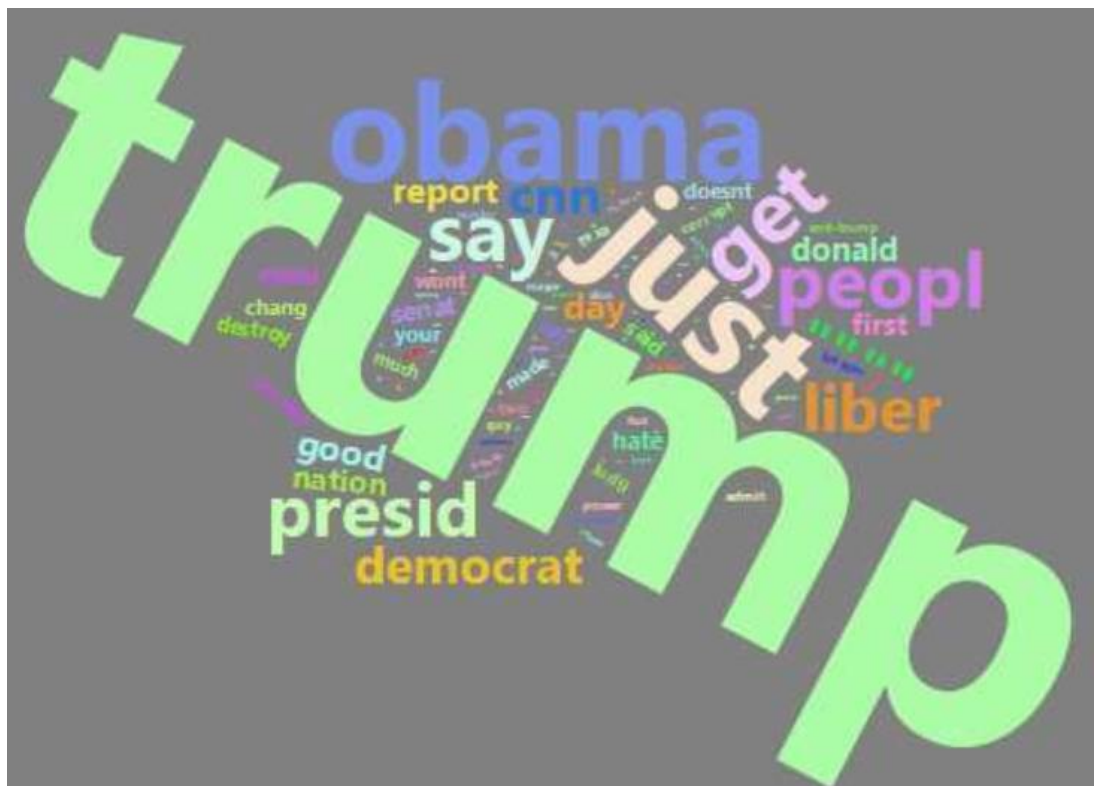

Image 2: Right Troll most frequent words

In order to get an idea of what content was in the Twitter data, word clouds were used to generate the most frequent words for each category. Image 1 contains the most frequent words for Left Troll. Upon examining the word cloud, issues involving black equality appear the most. Image 2 contains the most frequent words for Right Troll. Examining the word cloud reveals that the most frequent words are relating to President Donald Trump. Word clouds for Fearmonger, Hashtag Gamer and News Feed can be found in the appendix.

The Facebook data required a large amount of data preprocessing. The data needed to be extracted from each pdf and stored in a csv file. This process would be painstakingly slow to do manually, fortunately, this could be automated with the R package ‘pdftools’.  The structure for each pdf is similar to the one in Image A in the appendix. However, in the ‘Ad Targeting’ section, there are many different options and thus making extraction difficult. Therefore, Ad Targeting was dropped. Another issue was dealing with redacted information in the Ad Text. Redacted information showed up as a string of question marks. If the ad was fully redacted, it was dropped. In order to extract the user account, the name had to be stripped form ‘Ad Landing Page’. Ad Landing Page would often include links to sites other than Facebook or to an event. To look up all the ads would take too long, therefore there is an unknown and event category in the ‘account’ variable.

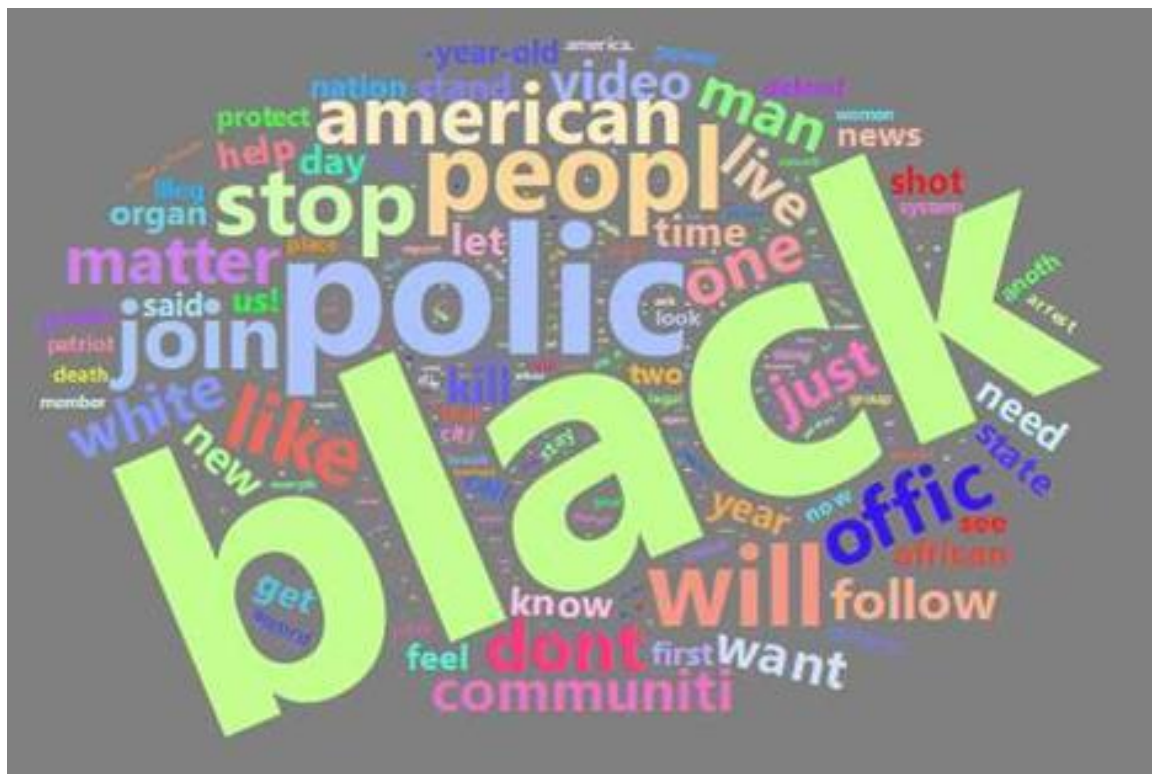

Image 3: Facebook ad most frequent words

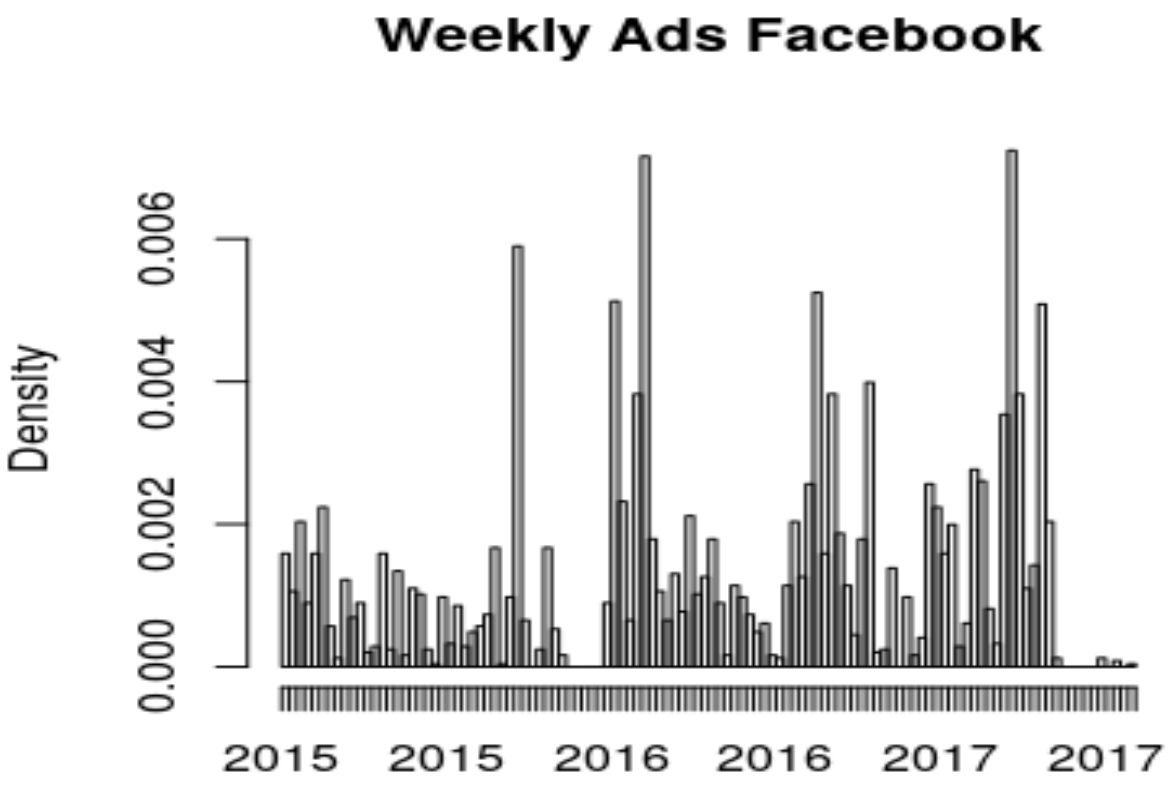

Figure 1: Weekly ads generated

To better understand the content in the Facebook ads, word cloud generated the most frequent words for all ads. The word cloud looks similar to Left Troll word cloud found in Image 1. Further exploration of the data seemed to indicate that ads where generated based on current news topics. During the time Facebook time frame, there were on going issues with police brutality and black equality. Therefore it is hypothesized that that there should be accounts falling into News Feed and Left Troll categories. Figure 1 gives the weekly ads generated for Facebook. The dates of the Facebook ads range from March 2015 to November 2017. It should be noted that the full time line of all ads is unknown. It is observed that there are gaps in 2016 and 2017. It is not believed that there were no ads generated in these periods. Rather, it is assumed that the ads were censored by the committee.

## Method

In this section, we will discuss several methods that we used for our project.

### LDA:

Latent Dirichlet allocation (LDA) is a generative probabilistic model for collections of discrete data such as text corpora. LDA is a three-level hierarchical Bayesian model, in which each item of a collection is modeled as a finite mixture over an underlying set of topics. Each topic is, in turn, modeled as an infinite mixture over an underlying set of topic probabilities. In the context of text modeling, the topic probabilities provide an explicit representation of a document. LDA does not care the order of the words in the document. Usually, LDA use the bag-of-words feature representation to represent a document. To understand LDA, we need to discuss some definitions and notations. A word is the basic unit of discrete data, defined to be an item from a vocabulary indexed by {1, …., V}. We represent words using unit-basis vectors that have a single component equal to one and all other components equal to zero. A document is a sequence of N words denoted by w = ($w_1$, $w_2$, ..., $w_N$), where $w_n$ is the nth word in the sequence. A corpus is a collection of M documents denoted by D = {$w_1$, $w_2$, ..., $w_M$}.

Then to find a probabilistic model of a corpus that not only assigns high probability to members of the corpus, but also assigns high probability to other "similar" documents.

LDA assumes the following generative process for each document w in a corpus D:

1. Choose N ∼ Poisson(ξ).

2. Choose θ ∼ Dir(α).

. For each of the N words $w_n$:

(a) Choose a topic $z_n$ ∼ Multinomial(θ).

(b) Choose a word $w_n$ from $P(w_n | z_n, \beta)$, a multinomial probability conditioned on the topic $z_n$.

**Random forest**

Random forest is a tree-based algorithm which involves building several trees (decision trees), then combining their output to improve generalization of the model. Tree actually stratifies the data into distinct non-overlapping regions based on rules that determine a variables contribution to the homogeneity of the child nodes. It determines the highest homogeneity in child nodes to become root node. A split will occur if the Gini index for a child node is less than the parent node. A Gini index defined by

$$G = \sum_{k=1}^{K} \hat{p}_{mk} (1 - \hat{p}_{mk})$$

Here $\hat{p}_{mk}$ represents the proportion of training observations in the *m*th region that are from the *k*th class. Variables are also split on Entropy which measures node impurity. Entropy is a measure of node impurity. For a binary class (a,b),

$$\text{Entropy} = -p(a) * \log\big(p(a)\big) - p(b) * \log\,(p(b))$$

Entropy is maximized at p = 0.5 and minimized at p = 0 or 1.

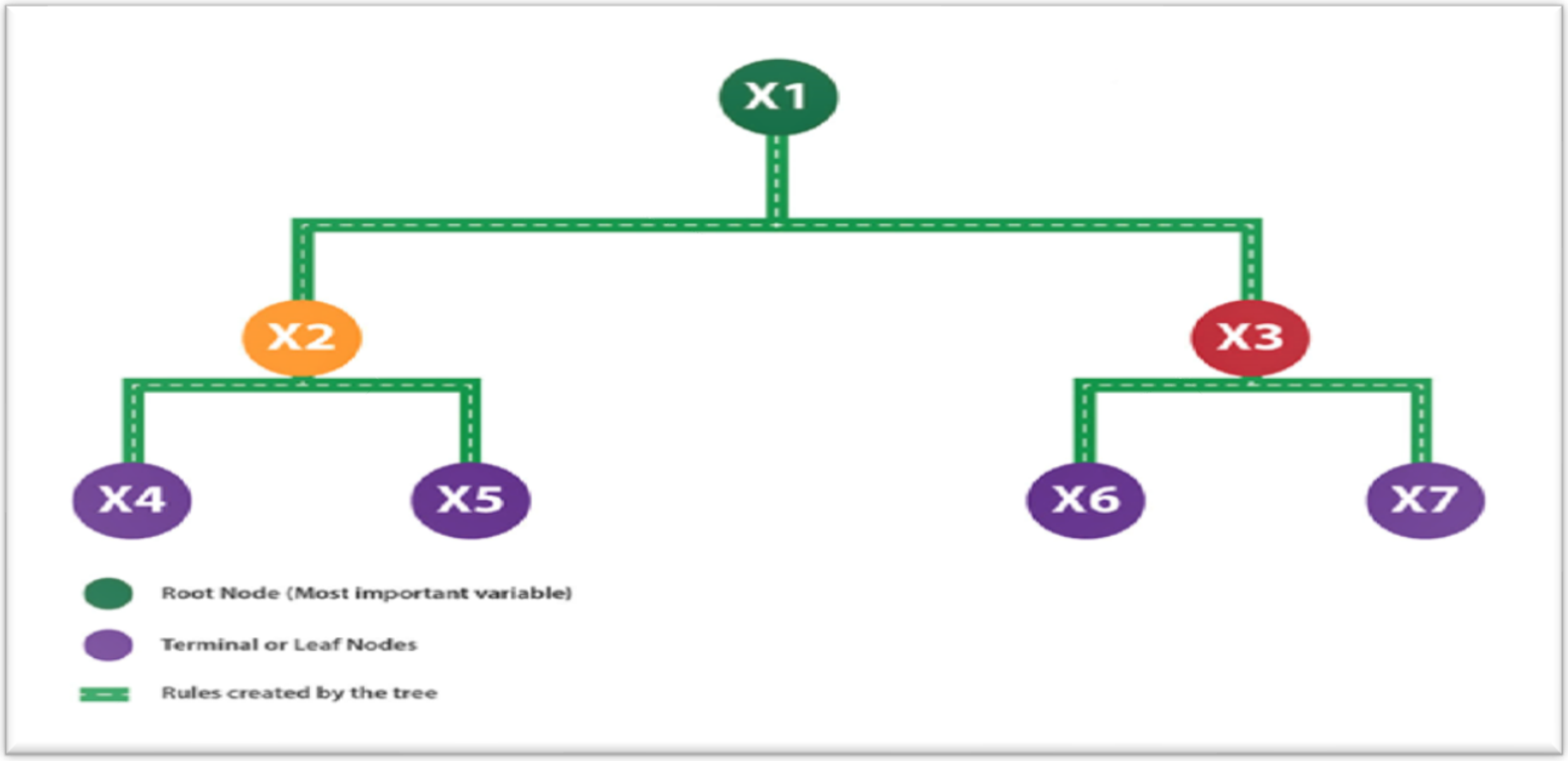

In the image above, the variable X1 resulted in highest homogeneity in child nodes, hence it became the root node. A variable at root node is also seen as the most important variable in the data set. X2 and X3 are the child nodes and rest of them are leaf nodes.

**SVM**

Support Vector Machine (SVM) is a supervised machine learning algorithm for classification and regression. It implements a technique called the kernel trick to transform the data. Kernel could be linear, radial, or polynomial. Using the kernel SVM identify the optimal separating hyperplane which maximizes the margin of the training data. In two dimensions a hyperplane is flat one-dimensional subspace-in other words, a line. In three dimensions, a hyperplane is a flat two-dimensional subspace-that is, a plane. There may exist multiple separating hyperplanes, SVM will choose the hyperplane that is as far from the data of each category.

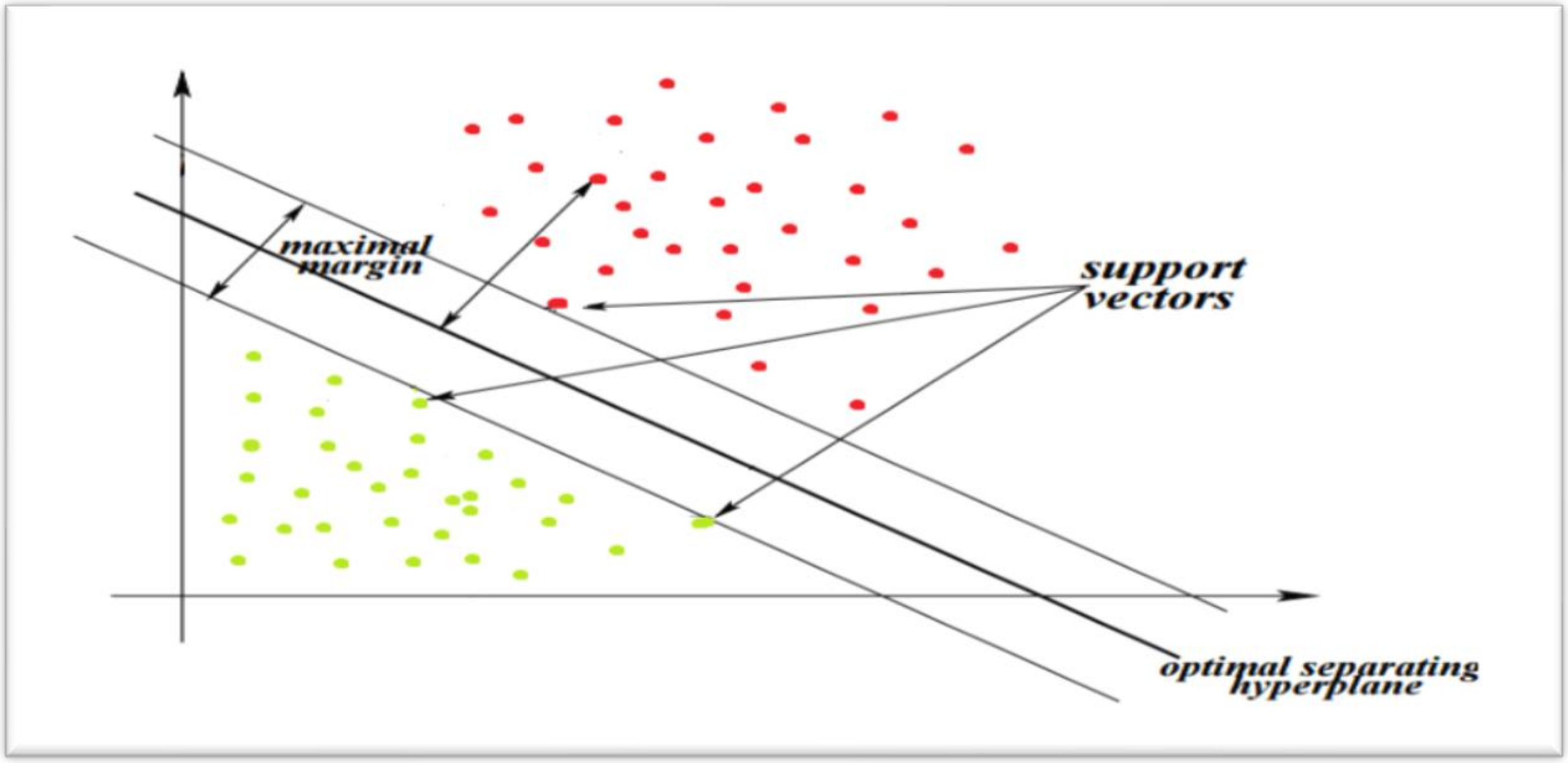

There are two classes of observations in the above figure, shown in green and in red. The maximal margin hyperplane is shown as a solid line. The margin is the distance from the solid line to either of the dashed lines. The two green points and the red point that lie on the dashed lines are the support vectors, and the distance from those points to the hyperplane is indicated by arrows. The red and green grid indicates the decision rule made by SVM based on this separating hyperplane.

**Results**

In order to classify the data, a Support Vector Machine classifier algorithm was used. A stratified sample of 250,000 tweets were generated from the Twitter data set to be used to train the model. A test set of 53,000 tweets was also generated to judge the performance of the model. Table 1 gives the truth table for the classification of the test data.

| SVM Truth Table | | | | | |
|---|---|---|---|---|---|
| Predict/Truth | Fearmonger | Gamer | Left Troll | News Feed | Right Troll |
| Fearmonger | 72 | 6 | 7 | 0 | 17 |
| Gamer | 2 | 742 | 85 | 39 | 95 |
| Left Troll | 36 | 1250 | 4562 | 523 | 1641 |
| News Feed | 12 | 505 | 1497 | 13092 | 2243 |
| Right Troll | 79 | 2681 | 5609 | 2881 | 15532 |

Table 1

The table is read from left to right where each column gives the count of classifications in the category. The diagonal of the table gives the number of correctly classified observations in the test set. Table 2 gives the percent of correctly classified observations for each category.

| SVM Prediction Accuracy | | | | |
|---|---|---|---|---|
| Fearmonger | Gamer | Left Troll | News Feed | Right Toll |
| 70.6% | 77.1% | 56.9% | 75.5% | 58% |

Table 2

From examining both tables, it is obvious to see that Fearmonger, Hashtag Gamer and News Feed are easily separated by the algorithm. However, Left Troll and Right Troll have issues being separated. To better separate them, other kernels were looked at, such as radial and polynomial. These kernels did poorly with radial classifying nearly 100% of accounts as Right troll and polynomial classifying 90% of accounts as Fearmonger. It is theorized that Left and Right troll have correlated words between them since both are talking heavily about political issues that pertain to both sides. Therefore the correlation creates issues for SVM to separate the two categories.

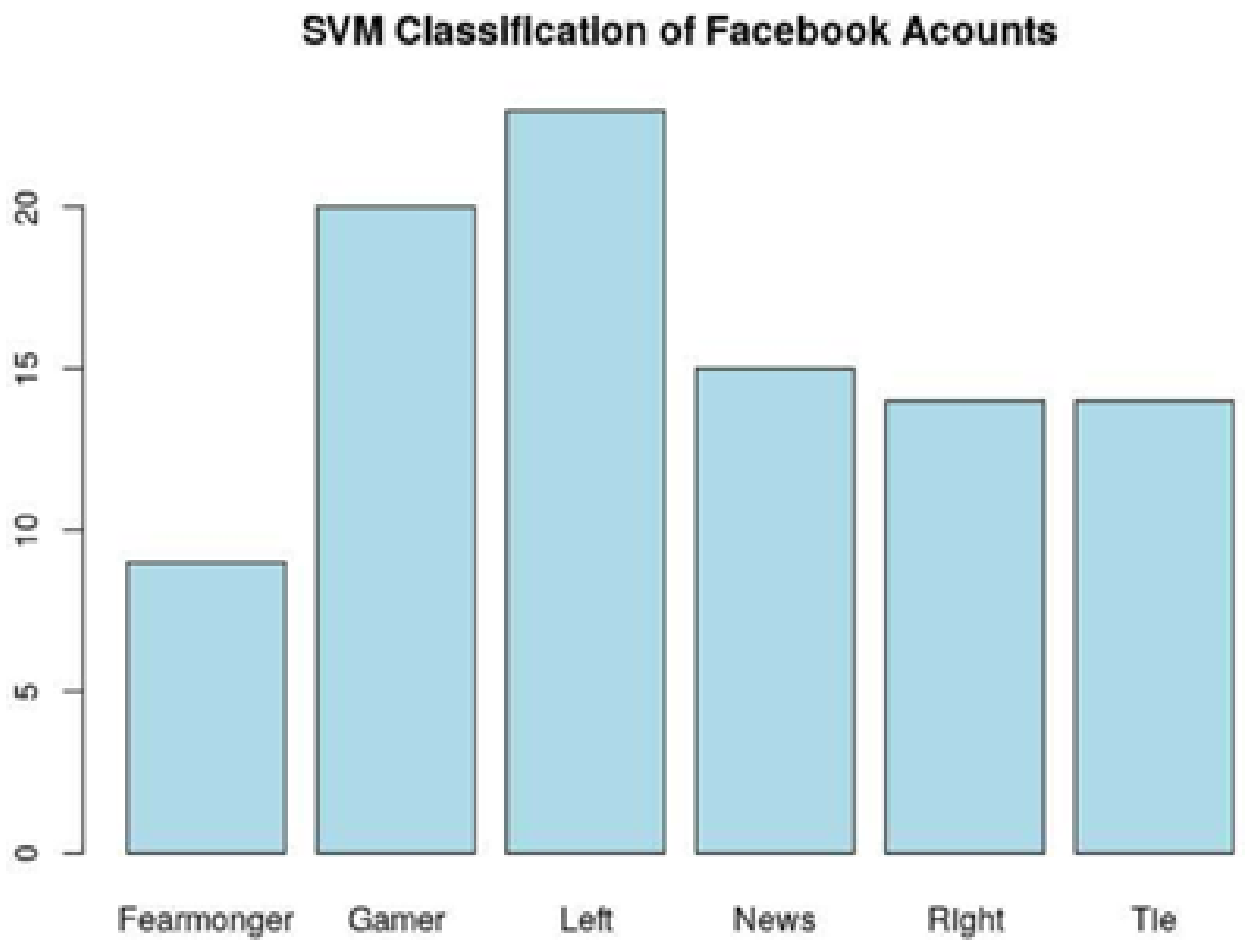

Figure 2: Counts of accounts classified

From the above model, the data on the 95 identified Facebook accounts were fed in. Figure 2 gives the counts for the number of accounts in each category. There is an extra category, ‘Tie’, for accounts who had a tie between more than one categories. The figure shows a contradiction to the hypothesis that News Feed and Left Troll would have the most classified into them. Left troll does have the highest count, however, Hashtag Gamer was close to it. There were 14 ties and most of these are associated with accounts that have few ads, usually less than five total. These may need to be classified by a human.

Since there were issues of separating Left Troll and Right Troll, topic models were generated from the data set. Topic models are topics that are generated from a set of documents and the topics are naturally occurring across documents. The natural occurring topics can be compared to the categories generated by Linvill and Warren. Table 3 shows the result of 6 topics that were generated by LDA.

| Topic 1 | Topic 2 | Topic 3 | Topic 4 | Topic 5 | Topic 6 |
|---|---|---|---|---|---|
| photo | music | star | blacklivesmatt | fuck | topvideo |
| happi | beat | honor | target | lost | futur |
| midnight | check | season | fall | shit | rise |
| dog | listen | red | bank | human | interest |
| babi | book | movi | kansa | TRUE | rate |
| celebr | 2017 | entertain | battl | what | travel |
| littl | morn | jame | road | yes | launch |
| buy | readi | brown | drive | hillaryclinton | grow |
| mom | retweet | footbal | hurt | hell | strong |
| heart | youtub | film | burn | wonder | land |
| christma | nowplay | paul | storm | mind | reach |
| food | art | coach | shut | shes | challeng |
| beauti | tonight | antitrump | store | phone | educ |
| walk | weekend | michael | updat | perfect | propos |
| eat | host | hire | arizona | idea | posit |
| parent | king | mark | pull | trust | trade |
| wear | artist | award | church | mad | israel |
| wow | latest | resign | respond | import | market |
| eye | song | loss | seattl | respect | plane |
| brother | rock | bush | hill | ass | prepar |
| amaz | chanc | mike | danger | dream | massiv |
| daughter | sell | 2015 | thousand | moment | emerg |
| light | facebook | hollywood | wisconsin | damn | result |
| bless | sale | detroit | flood | forget | huge |
| surpris | super | hero | rob | stupid | camp |
| wish | friday | ryan | file | explain | price |
| vega | track | michigan | steal | sound | carri |
| pictur | app | memori | beach | worst | doubl |
| drink | interview | scandal | due | wakeupamerica | fed |
| cat | messag | director | rescu | hey | key |

Table 3: LDA Topics in Twitter Data

From Table 3, one can observe the first 30 words of each topic. The first word has the highest probability of being in that topic. After examining the topics and the words in them, the following topics were assigned formal labels. Table 4 gives the labels that were thought to best describe the category.

| Topic 1 | Topic 2 | Topic 3 | Topic 4 | Topic 5 | Topic 6 |
|---|---|---|---|---|---|
| Family | Trending | Pop Culture | Negative News | Anti-Hillary | News |

Table 4

LDA was specifically set to generate six topics in order to match the six categories in the Twitter set. The optimal k topics for the set was computed to be 8. After examining the 8 topics, most topics could not be properly labeled due to no overall theme. Therefore it was concluded that the optimal k topics was not useful. From Table 4, the labels that best summarize the topics are not exactly representative of the categories that were generated by Linvill and Warren. It could be argued that Negative News and News could be representative of the News Feed category. However, LDA has determined that these are two separate topics not one topic.

| Topic 1 | Topic 2 | Topic 3 | Topic 4 | Topic 5 | Topic 6 |
|---|---|---|---|---|---|
| join | blackmattersus | communiti | right | like | free |
| matter | love | support | unit | know | need |
| life | beauti | gun | nation | veteran | music |
| news | kid | racism | stand | say | trump |
| togeth | african | student | illeg | good | world |
| bring | children | 2nd | never | care | facemus |
| event | woman | patriot | citi | follow | home |
| selfdefens | women | great | immigr | clinton | believ |
| class | proud | defend | protect | hillari | share |
| safe | girl | amend | hate | texa | click |
| free | look | social | brutal | issu | muslim |
| brutal | usa | tell | group | real | onlin |
| donat | teen | busi | flag | problem | musicfbinfo |
| king | histori | univers | member | truth | peac |
| stay | power | lover | equal | wont | refuge |
| parti | melanin | offici | south | rememb | job |
| freedom | child | rais | cant | money | player |
| prison | goal | thank | better | noth | repost |
| basi | word | injustic | border | heart | keep |
| break | cultur | teacher | part | thing | browser |
| ralli | parent | chicago | discrimin | without | brother |
| sure | unite4justic | studi | understand | lie | legal |
| train | best | voic | civil | old | listen |
| racial | woke | move | god | subscrib | much |
| miss | got | develop | transgend | media | find |
| latest | middl | institut | everyon | talk | jump |
| anniversari | viral | hes | church | tamir | learn |
| jerom | know | inspir | movement | guy | download |
| privileg | queen | marin | speak | turn | secur |
| african | forget | team | lgbt | bad | absolut |

Table 5

The same procedure was done with the Facebook data. It is believed that the Twitter categories will not align with the topics generated by LDA with the Facebook data. Again, LDA is set to produce six topics to match the six account categories. Table 5 gives the results for the Facebook data. Table 6 gives the Labels that were best representative of the topics generated. From Table 6, Topic 1 and Topic 2 are closely related to each other and similar to what one might find in the Left Troll category. Also, Topic 3 and Topic 4 are similar to what one might find in the Right Troll category. There is also a topic that had no overall pattern of within it. This was classified as 'Other'. The last topic was labeled as social media due to the advertisement like words one would find on social media. This topic was similar to the 'Trending' topic in the Twitter data. The presence of these two topics is not surprising due to the fact the content is from social media platforms.

The optimal k topics was also generated and found to be 10 topics. This was necessary to run due to the fact that the Facebook data may contain more or less topics than the Twitter data. However, after examining the ten topics, no appropriate labels could be determined for the topics. The ten topics can be found in the Appendix under Table A.

| Topic 1 | Topic 2 | Topic 3 | Topic 4 | Topic 5 | Topic 6 |
|---|---|---|---|---|---|
| Black Matters | Black Empowerment | Anti-$2^{nd}$ Amendment | Anti-immigration | Other | Social Media |

Table 6

**Discussion**

The initial goal of this analysis was to examine if the Twitter categories existed in the Facebook data. To answer this question, two algorithms were to be used to examine and evaluate this question. Random Forest and Support Vector Machines were selected due to their ability to classify multiple categories simultaneously. However, due to the nature and size of the data, personal computers could not accommodate both the memory and the computational requirements needed to run these algorithms. To combat this this issue, the data and code were imported on to a computing cluster. The cluster was able to run all the code efficiently. However, in order to speed up the Random Forest algorithm, it needed to be run in parallel. This could not be achieved due to an issue with R calling fork() with openMPI. A parent process will call fork to create a copy of itself to spawn a child process. In the newest version of openMPI, the program throws a fatal error and kills the program. This is not an issue that can be fixed and caused the Random Forest to be excluded from the results.

Although there was issues with Random Forest, SVM produced satisfactory results. One way to improve the results is using the one-verses-all method. This would group all categories together while leaving one out. This can produce a stronger fit, however this proved to be too involved with the given time constraint. This should be investigated later.

## Conclusion

Given only a representative sample of the Facebook ads created by the IRA, it is unknown if the Twitter categories exist in the entire Facebook set. For the sample, it is not believed that the categories exist, at least not all of them. Further analysis and better techniques should be used to further investigate this question. It is believed that neural networks could give better classifications. The analysis has shed light into how the IRA used both platforms. The IRA used news events to create advertisements to capture enraged citizens and spread their propaganda and disinformation. Twitter was used to distract and divide the country in an attempt to undermine democracy.

## Appendix

Image A:

| | |
|---|---|
| Ad Text | Every man should stand for our borders! Join! |
| Ad Landing Page | https://www.facebook.com/Secured-Borders-757183957716200/ |
| Ad Targeting | Location: United States: Alabama; Georgia; New Mexico; Arizona (+50 mi) Texas<br>Excluded Connections: Exclude people who like Secured Borders<br>Age: 17 - 65+<br>Gender: Male<br>Placements: News Feed on desktop computers or News Feed on mobile devices<br>People Who Match: Interests: Right to keep and bear arms, The Second Amendment, Stop Illegal Immigration, National Rifle Association or Donald Trump for President |
| Ad Impressions | 2,502 |
| Ad Clicks | 258 |
| Ad Spend | 3,025.52 RUB |
| Ad Creation Date | 10/11/16 09:11:47 AM PDT |
| Ad Start Date | 02/09/16 03:34:26 AM PST |

Figure A:

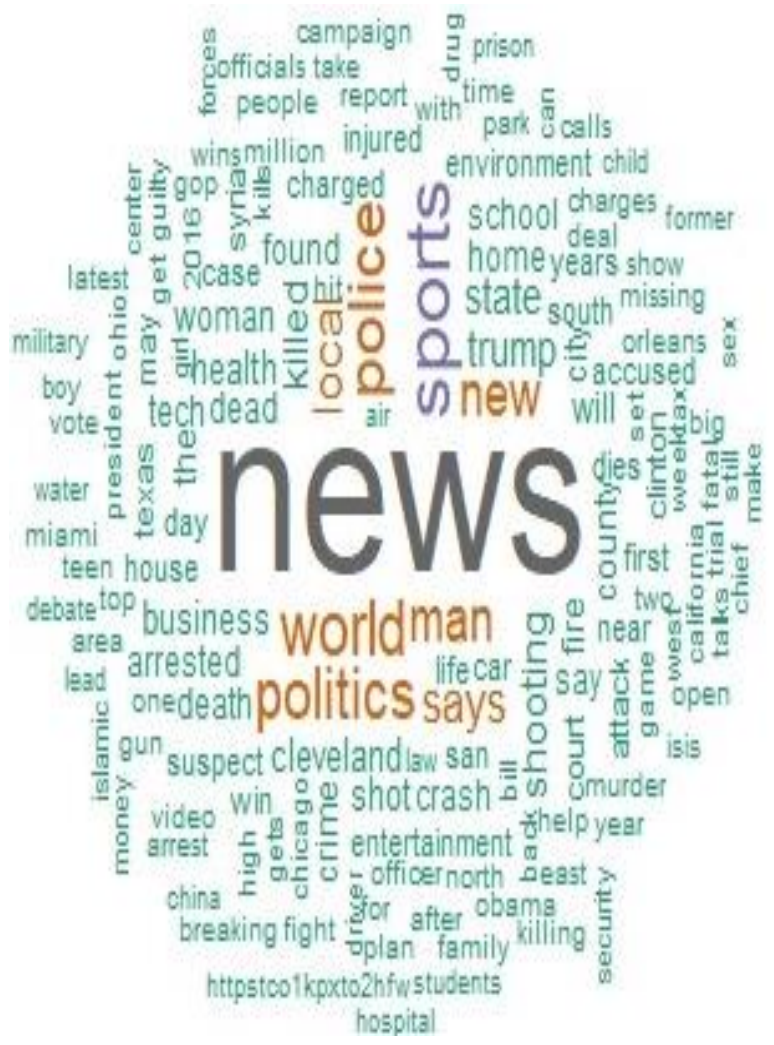

Word Cloud for News Feed

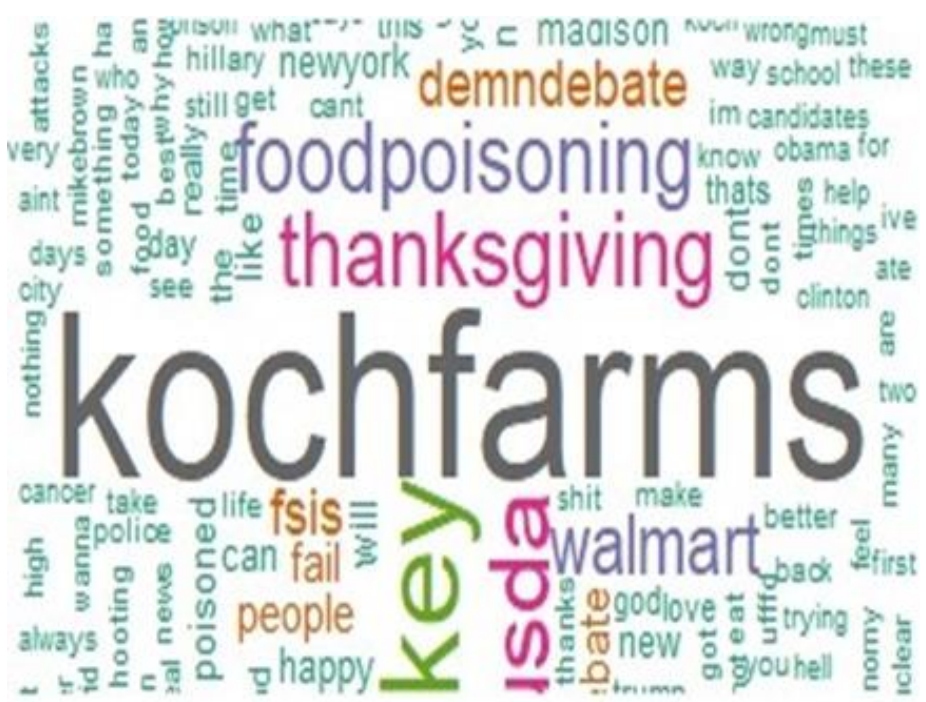

Word Cloud for Fearmonger

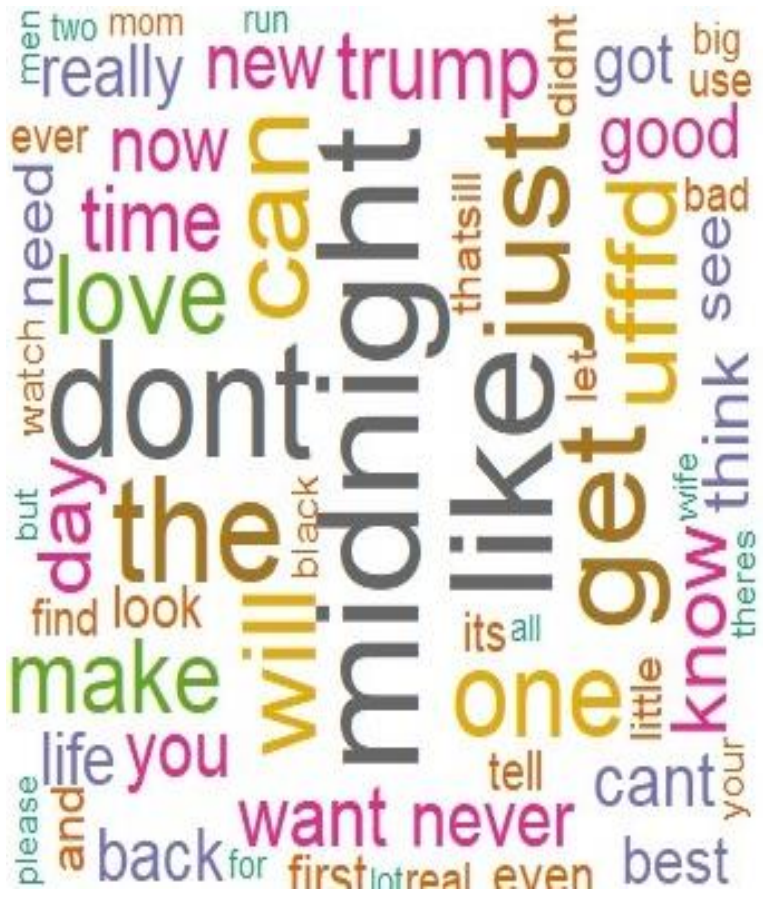

Word Cloud Hashtag Gamer

Table A

| Topic 1 | Topic 2 | Topic 3 | Topic 4 | Topic 5 | Topic 6 | Topic 7 | Topic 8 | Topic 9 | Topic 10 |
|---|---|---|---|---|---|---|---|---|---|
| join | need | brutal | matter | communiti | unit | like | love | free | right |
| free | trump | racism | african | support | veteran | illeg | beauti | music | life |
| bring | peac | ackmattersusco | histori | gun | muslim | say | know | world | togeth |
| event | believ | student | teen | news | hate | good | woman | facemus | home |
| selfdefens | refuge | stand | flag | 2nd | clinton | immigr | women | onlin | share |
| class | care | citi | girl | patriot | hillari | texa | power | musicfbinfo | children |
| safe | thing | kid | south | defend | learn | parti | usa | click | keep |
| proud | that | nation | ever | amend | social | border | melanin | player | equal |
| donat | job | never | ralli | follow | issu | part | goal | browser | king |
| stay | problem | cant | break | busi | group | old | transgend | repost | child |
| basi | wont | univers | thank | lover | nation | without | unite4justic | legal | like |
| sure | real | discrimin | tell | heart | member | talk | got | listen | freedom |
| train | money | offici | confeder | rais | understand | tamir | woke | brother | prison |
| miss | protect | everyon | racial | photo | church | look | nativ | jump | great |
| april | media | movement | chicago | latest | god | protect | queen | everyth | find |
| enjoy | rememb | word | race | loui | speak | feder | cultur | download | voic |
| feb | secur | meet | great | develop | truth | full | look | welcom | civil |
| buy | syria | teacher | studi | tea | lgbt | rice | weapon | absolut | parent |
| coach | end | middl | noth | key | christian | lie | land | never | better |
| meetup | hero | viral | rape | best | terror | anniversari | much | keep | case |
| save | leav | garner | subscrib | brave | polit | beat | divers | mexican | turn |
| supremacist | march | eric | inspir | camera | play | hell | femal | add | watch |
| brown | forget | facebook | jone | liberti | across | move | mental | eye | boy |
| orlando | futur | realiz | heritag | opinion | armi | three | dope | sister | privileg |
| resist | bomb | institut | guy | skin | islam | bad | lesbian | gray | jerom |
| tampa | came | know | misti | everyday | jesus | unit | davi | hair | mom |
| care | charlott | racial | aiyana | instagram | east | hall | pass | via | dream |
| mission | subscrib | injustic | water | jason | hes | respect | natur | freddi | count |
| betti | what | marin | william | account | deserv | page | girl | extens | drive |
| prioriti | becam | abus | symbol | meme | father | star | perform | blind | percent |